\documentclass[a4paper,11pt]{article}
\pdfoutput=1 % if your are submitting a pdflatex (i.e. if you have
\usepackage{jheppub} % for details on the use of the package, please
\usepackage[T1]{fontenc} % if needed
\usepackage{graphicx}% Include figure files
\usepackage{dcolumn}% Align table columns on decimal point
\usepackage{bm}% bold math
\usepackage{hyperref}% add hypertext capabilities
\usepackage{tabularx}
\usepackage{multirow}
\usepackage[mathlines]{lineno}% Enable numbering of text and display math
\title{\boldmath Photon detection probability prediction \\ using one-dimensional generative neural network}

\author[1]{Wei Mu\note{Corresponding author.},}
\author{Alexander I. Himmel, }
\author{and Bryan Ramson}

\affiliation{Neutrino Division, Fermi National Accelerator Laboratory, \\Wilson Street and Kirk Road, Batavia, Illinois, 60510 U.S.A.}

\emailAdd{wmu@fnal.gov}

\abstract{Photon detection is important for liquid argon detectors for direct dark matter searches or neutrino property measurements. 
Precise simulation of photon transport is widely used to understand the probability of photon detection in liquid argon detectors. Traditional photon transport simulation, which tracks every photon using the \textsc{Geant4} simulation toolkit, is a major computational challenge for kilo-tonne-scale liquid argon detectors and GeV-level energy depositions. 
In this work, we propose a one-dimensional generative model which efficiently generates features using an $\mathrm{OuterProduct}$-layer. 
This model bypasses photon transport simulation and predicts the number of photons detected by particular photon detectors at the same level of detail as the \textsc{Geant4} simulation. 
The application to simulating photon detection systems in kilo-tonne-scale liquid argon detectors demonstrates this novel generative model is able to reproduce \textsc{Geant4} simulation with good accuracy and 20 to 50 times faster.
This generative model can be used to quickly predict photon detection probability in huge liquid argon detectors like ProtoDUNE or DUNE.}

\begin{document} 
\maketitle
\flushbottom

\section{\label{sec:intro}Introduction}

Liquid argon (LAr) is a popular detector medium for direct dark matter searches \cite{calvo2017commissioning, aalseth2018darkside, benetti2008first} and neutrino property measurements \cite{abi2020first, bettini1991icarus, acciarri2017design}. 
Its excellent scintillation properties have been employed by dark matter search experiments for energy reconstruction and background rejection.
LAr scintillation light detection will advance the physics goals of neutrino experiments by improving detector calorimetric and position resolution, enabling the study of astrophysical neutrinos and enhancing the physics reach of oscillation analyses. 
In order to take the advantage of information provided by LAr scintillation light, photon detection probability in LAr detectors must be well understood.

Traditionally, photon detection probability is explored by simulating the transport of photons in detectors using \textsc{Geant4} \cite{agostinelli2003geant4}.
However, such simulation is extremely challenging for experiments using huge LAr detectors that record GeV-level energy depositions due to limited computing resources. 
Modern machine learning techniques have enabled new ways to emulate the results from full \textsc{Geant4} simulation and a generative model is one of the most promising approaches for learning the true distribution from training samples so as to generate new data points with variation~\cite{Alonso-Monsalve:2018aqs}. 
However, while generative models based on deep neural networks (\textsc{DNN}) have shown great promise in generating accurate predictions, they can be too slow when deployed without GPUs, motivating the development of a novel generative model which can efficiently predict photon detection probabilities at lower computational cost.
%inference must be run without GPUs,
%Nowadays, two popular solutions have been implemented to quickly emulate results from full \textsc{Geant4} simulation: tabulating photon detection probabilities with a photon library or parameterizing the effect of photon transport using semi-analytical model. The photon library method relies on a lookup table which gives the probabilities that photons produced in particular positions are observed by particular photon detectors \cite{abi2020deep}. The semi-analytical model uses the detector geometry to predict the amount of light detected, and employs parameterized corrections to this geometric prediction to account for second order transport effects  \cite{garcia2021predicting}. While both methods produce valuable results, there are also limitations: the photon library method is difficult to scale up for huge LAr detector because the lookup table will be too large to be held in memory, while the semi-analytical model requires specific knowledge and human efforts to tune multiple parameters. Therefore, new approaches to efficiently predict photon detection probabilities are needed.

In this paper, we demonstrate that a one-dimensional generative neural network (1D \textsc{Genn}) is capable of bypassing photon transport simulation and rapidly predicting precise photon detection probabilities based only on the scintillation vertex. 
We first discuss common features of photon detection systems for LAr detectors and propose a general architecture for the \textsc{Genn} in Section~\ref{sec:network}. 
Subsequently, we apply and train \textsc{Genn} models for photon detection systems of two large scale LAr detectors. 
In Section~\ref{sec:perf}, we evaluate the performance of the \textsc{Genn}-based photon detection probability predictions, in comparison to full \textsc{Geant4} simulations, on a number of metrics including: the capability to smoothly interpolate, the precision of the prediction, and the speed of inference on CPUs. 
In addition, we evaluate the scalability of the \textsc{Genn} model to both higher precision and to larger systems. 
Finally, we summarize our work and discuss the potential generalization of the \textsc{Genn} model in Section~\ref{sec:con}. 

\section{\label{sec:network}1D generative model}

Aiming at building a fast and precise generative model for photon detection probability prediction, we extract common features from photon detection systems of LAr detectors, propose a novel network architecture that is able to conditionally generate the photon detection probabilities based on the scintillation vertex, and instantiate the model for two large scale LAr detectors.

\subsection{\label{subsec:arch}{Model architecture}}

Common photon detection systems of LAr detectors consist of a series of photon detectors distributed in arrays and deployed on the ``walls'' of the detector. 
Photons, emitted from a vertex where a particle deposits energy illuminate specific photon detectors and form a hit pattern on the photon detection system. 
The number of total detected photons, being correlated to the deposited energy of the particle, can be used as a calorimetric energy measurement and improve the detector energy resolution. 
The distribution of the illuminated photon detectors, which depends on the location of the scintillation vertex, can be used to locate the two-dimensional (2D) position of the interaction point. 
The \textsc{Geant4} toolkit uses a Monte-Carlo (MC) method to simulate photon transport inside a detector and produce the number of photons detected by particular photon detectors, which predicts the photon detection probability of the detector. 
As this MC-based method is challenging for huge LAr detectors and GeV-level energy depositions, the prediction from the full \textsc{Geant4} simulation might be emulated, in a more efficient way, by generative models, which directly map the energy deposition, with specific amount and position, to the hit pattern on the photon detection system.

Recently, generative models using \textsc{DNN} have shown the ability to randomly produce new high resolution images that mimic the details of true images. 
However, \textsc{DNN} inference on CPUs is slow. 
Millions of simulated events with billions of hit pattern images are required for physics analyses, suggesting significant computing resources.
Given that modern LAr detectors will deploy thousands of photon detectors and the photon detection systems of LAr detectors will generate hit pattern images consisting of only thousands of pixels, it could be a fairly costly and time consuming process which is far less than the images processed by common generative models. 
In addition, the images to be generated strongly depend on the location of the scintillation vertex, which makes it possible to fill in features to an image in a more efficient way. This motivated us to explore whether generative models based on shallow neural networks with novel architectures could reproduce the features of the photon detection systems of LAr detectors.

Considering a LAr detector in a three-dimensional (3D) Cartesian coordinate system, a photon detector array is deployed perpendicular to the $x$-axis, where photon detectors are aligned in an $m \times n$ array along the $y$-axis and $z$-axis. 
Assuming a certain number of photons are emitted from a scintillation vertex ($x_{\rm s}, y_{\rm s}, z_{\rm s}$), it is a reasonable intuition that the number of detected photons by particular photon detectors depends on the coordinate $x_{\rm s}$ and the distribution of the illuminated photon detectors along row or column in the $yz$-plane is determined by the coordinate $y_{\rm s}$ or $z_{\rm s}$ .

\begin{figure}[htp]
\centering 
\includegraphics[width=0.6\textwidth]{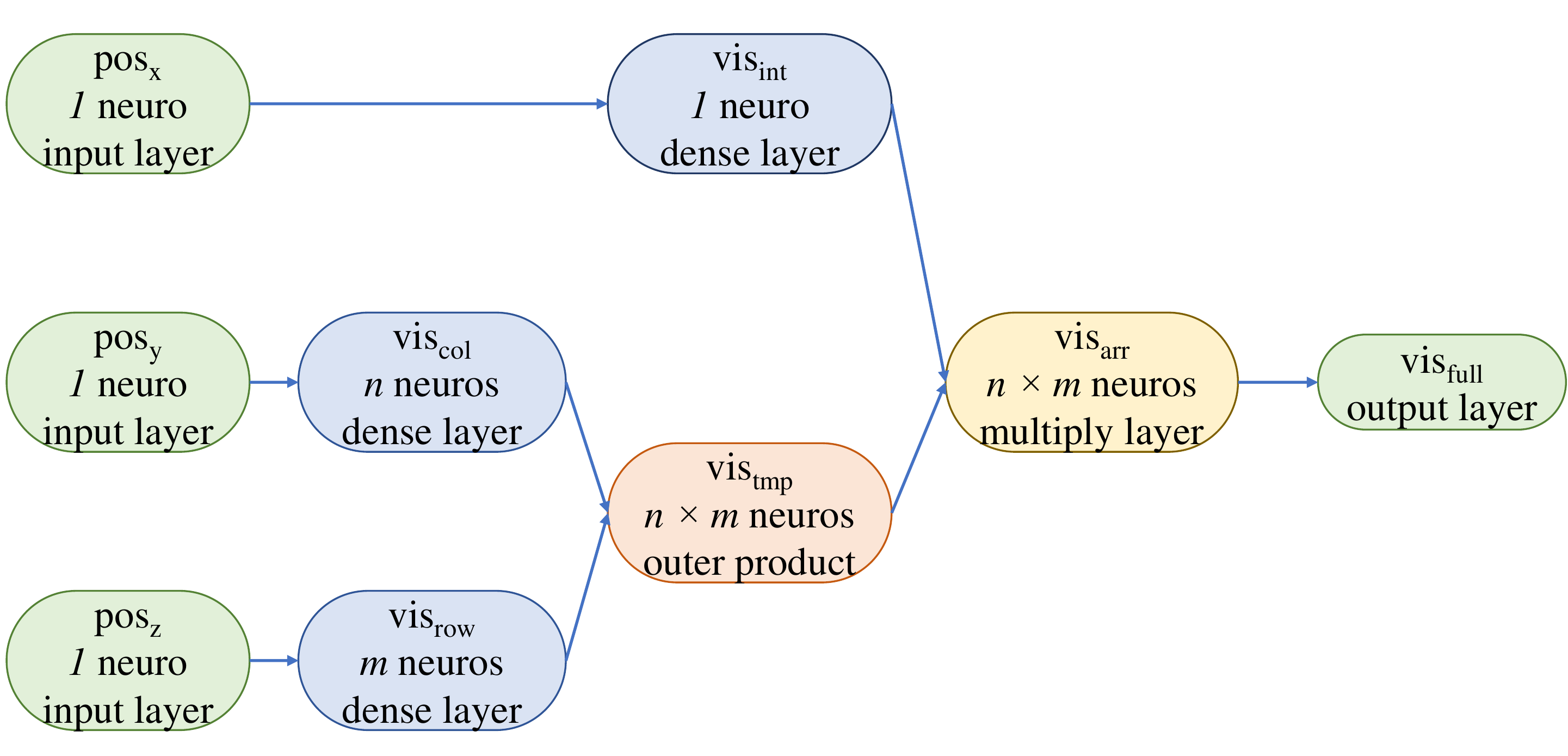} 
\caption{General \textsc{Genn} architecture. ${pos_y}$-layer and ${pos_z}$-layer are used to predict the photon distribution on the photon detection system, while ${pos_x}$-layer is a normalization factor for the number of total detected photons. $\mathrm{OuterProduct}$ layer expands the dimension of the inputs.} 
\label{fig:gnnarch}
\end{figure}

According to the common photon detection system design, we propose a general \textsc{Genn} architecture with three input layers, named ${pos_x}$, ${pos_y}$, and ${pos_z}$, each having one neuron corresponding to the coordinate of the scintillation vertex, $x_{\rm s}$, $y_{\rm s}$ , and $z_{\rm s}$. 
The \textsc{Genn} has one output layer, named ${vis_{\rm full}}$,  with number of neurons equal to the number of photon detectors. 
There are three first hidden layers, connecting to the three input layers respectively. The hidden layer connected to ${pos_x}$, named ${vis_{\rm int}}$, has one neuron as a factor to normalize the number of total detected photons. The hidden layer connected to ${pos_y}$  or ${pos_z}$, named ${vis_{\rm col}}$ or ${vis_{\rm row}}$, has $n$ or $m$ neurons, being used to predict the photon detection probabilities by detectors in each column or row.
The outer product of ${vis_{\rm col}}$ and ${vis_{\rm row}}$, composing an $\mathrm{OuterProduct}$-layer and named ${vis_{\rm tmp}}$, represents the distribution of the photons in the photon detector array. 
The multiplication of ${vis_{\rm int}}$ and ${vis_{\rm tmp}}$, forming a $\mathrm{Multiply}$-layer named ${vis_{\rm arr}}$, reflects the features of the hit pattern on the photon detector array and predicts the photon detection probability of each detector corresponding to a particular scintillation vertex.
In practice, the number of neurons on ${vis_{\rm col}}$ and ${vis_{\rm row}}$ ($n$ or $m$) can be tuned and optional layers can be introduced, such as Normalization layers, to get optimal network architecture (see Section~\ref{subsec:sca}). 
The general \textsc{Genn} architecture we proposed is illustrated in Figure~\ref{fig:gnnarch}.

Unlike common generative models, which use the upsample (UpSampling2D) layer or the transpose convolutional (Conv2DTranspose) layer for the input dimension expansion, the model proposed in this work uses a $\mathrm{OuterProduct}$ layer to produce features and expand the dimension of inputs, which significantly reduces the inference time on CPU. 
This model starts with a 3D position, passes 1D data from layer to layer, and finally gives rise to a prediction in a 1D data structure on the output layer. 
We will discuss the advantage of the 1D data structure in Section~\ref{subsec:loss}.

\begin{figure*}[htp]
\centering
\includegraphics[width=0.9\textwidth]{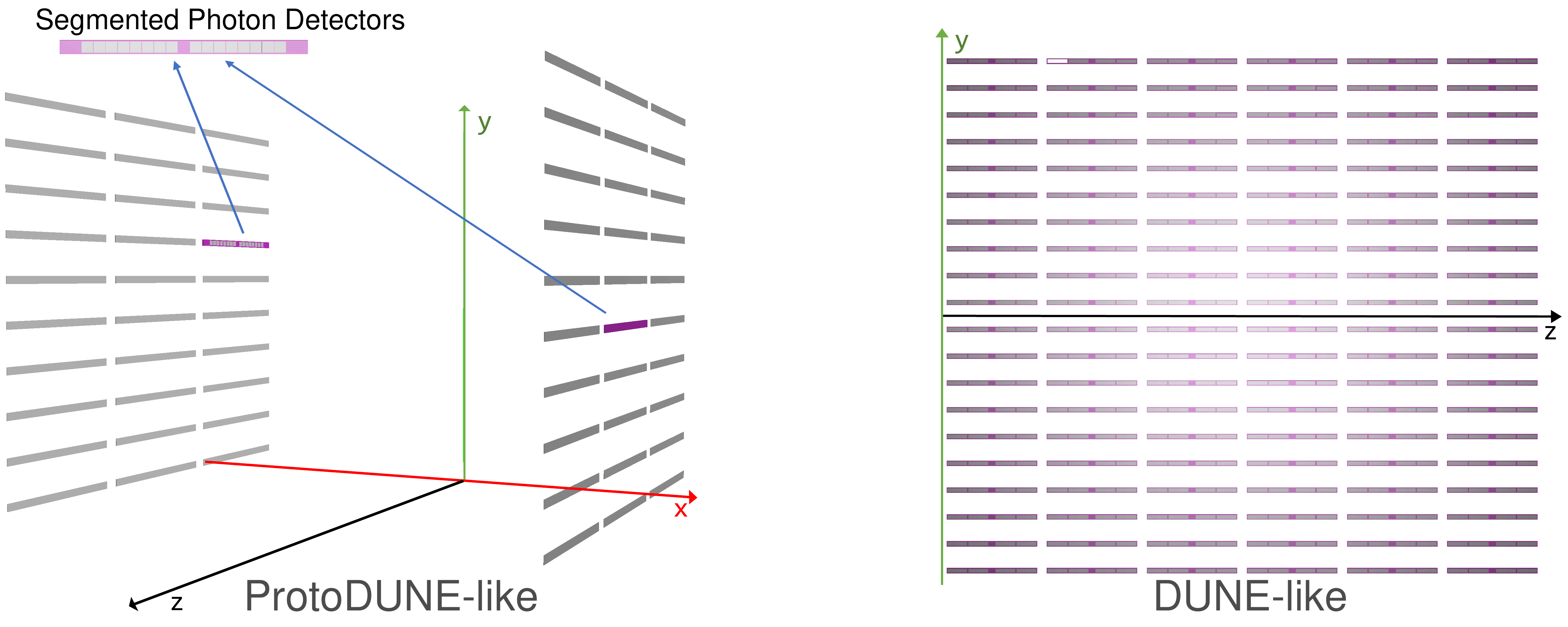}
\caption{Photon detection systems in ProtoDUNE-like (left) and DUNE-like geometries (right). The grey rectangles are the photon detectors.}
\label{fig:pddist}
\end{figure*}

This general \textsc{Genn} model has been applied to photon detection systems of ProtoDUNE-like \cite{abi2017single, abi2020first, dune2021design} and DUNE-like geometries \cite{abi2020deep}. 
Two photon detection arrays are deployed on two ``walls'' in ProtoDUNE-like geometry. 
In each array, two different types of photon detectors are used. 
30 bar-like light collectors are aligned in $3\times10$ array \cite{machado2016arapuca}, where one bar is segmented into 16 silicon photomultipliers, as showed in Figure~\ref{fig:pddist} on the left. 
Therefore, in each array, there are 29 bar-like and 16 segmented photon detectors. 
In total, 90 photon detectors are deployed in this geometry. 
The active volume sits between the photon detection systems, with an opaque wall halfway between them, so each wall detects the photons emitted on its side of the geometry.
In the DUNE-like geometry (a segment of a DUNE detector), 480 photon detectors are deployed on a single ``wall'' in the middle of the volume, forming a $yz$-plane. 
The photon detectors are rectangular and aligned with the $y$-direction forming 20 wide-spaced rows, each of which has 24 photon detectors placed end-to-end, as shown in Figure~\ref{fig:pddist} on the right. 
This photon detection system sits in the middle of the active volume, and so is able to detect photons emitted from both sides.

We instantiate 1D \textsc{Genn} models for above two photon detection systems. The networks have been implemented in the framework of \textsc{Keras}\footnote{\textsc{Keras}: https://keras.io/} on top of \textsc{TensorFlow}\footnote{\textsc{TensorFlow}: https://www.tensorflow.org/}, and the source code for the two instantiated models is available on GitHub\footnote{Source Code: https://github.com/Healthborn/genn}.

%The \textsc{Genn} model for ProtoDUNE photon detection system contains 18,831 (19,239) trainable (total) parameters while the one for DUNE geometry 145,292 (145,818) trainable (total) parameters

\subsection{\label{subsec:loss}Loss function}

Neural networks are generally trained using the gradient descent algorithm, which requires a loss-function to calculate model error and update model parameters using the back propagation algorithm. 
Common generative models generate 2D images, which makes it challenging to choose an optimal loss-function for model training since there is no consensus as to which measure best captures the feature difference between two 2D images. 
This motivates the generative model to be trained in the generative adversarial network (\textsc{Gan}) framework \cite{goodfellow2014generative, mirza2014conditional}, 
where a discriminator network is used to evaluate the output from the generative model. The \textsc{Gan} framework has led to success in many applications, but objective and quantitative evaluation of \textsc{Gan} generative model remains an open question. 
So far, there is no general agreement upon algorithms to find the Nash-equilibrium between the discriminator and the generative model, and it relies on human eyes to determine performance of the generative model, which results in an unstable training procedure. 
We solve this problem by training the \textsc{Genn} models with a clearly defined loss-function instead of within the \textsc{Gan} framework.

It is natural to assume generative models would learn features from 2D images in the similar way as human brain. 
However, neural networks might view data in a different way and learn features from flattened 2D images: even non-spatially-representative 1D-vectors.
In this case, the problem to be solved by generative models can be considered a multiple regression prediction problem, and the 1D data structure allows more straightforward similarity metrics \cite{santini1999similarity, goldberger2003efficient, gretton2012kernel}.

The most commonly used metric for a multiple regression problem, ${mean~squared~error}$ (MSE) or $mean~absolute~error$ (MAE), tends to produce ${regression~to~the~mean}$ situation, therefore, we design a loss-function based on physics performance. 
As the value of each element of the 1D-vector predicted by the \textsc{Genn} model is photon detection probability, the Kullback–Leibler (KL) divergence ($D_{\rm KL}$) provides a good starting point, which is defined as:
\begin{equation}
 D_{\rm KL} (P || Q) = \sum_{x} P(x) \log \frac{P(x)}{Q(x)} ,
\label{eq:kld}
\end{equation}
where $P(x)$ corresponds to the ``test'' distribution, and $Q(x)$ the ``true'' distribution. 
However, the KL-divergence has a problematic property that $D_{\rm KL}$ dislikes regions where $Q(x)$ has non-null value and $P(x)$ has null value, which is an excellent feature in certain situations but not good for this application. 
Also, because $D_{\rm KL}$ is value-weighted by the probability $P(x)$, the $D_{\rm KL}$-loss-function leads to a selection bias where the error measured by low-value elements is underrated.

Motivated by one of the variations of KL-divergence: Jensen–Shannon (JS) divergence, we propose a loss-function using a variational KL-divergence $D_{\rm vKL}$, defined as:
\begin{equation}
 D_{\rm vKL} (P || Q) = \bigg |\sum_{x} \Big ( P(x)-Q(x) \Big ) \log \frac{P(x)}{Q(x)} \bigg | ,
\label{eq:vkld}
\end{equation}
where $x$ stands for the sequence number of photon detectors, $P(x)$ the ``test'' 1D-vector from \textsc{Genn} model prediction, and $Q(x)$ the ``true'' 1D-vector from \textsc{Geant4} simulation.
The $D_{\rm vKL}$-loss-function is symmetric and well behaved no matter whether either $P(x)$ or $Q(x)$ is small or not. 
Because $D_{\rm vKL}$ is value-weighted by the difference between $P(x)$ and $Q(x)$, it guarantees the model to converge to the optima. %\TK{Would be nice to explain a bit more here why this loss function has better properties if there is a straightforward a priori explanation.}
The training of sample \textsc{Genn} models using the $D_{\rm vKL}$-loss-function proves $D_{\rm vKL}$ successfully captures properties of this problem. %This $D_{\rm vKL}$-loss-function can be generalized for any 1D generative model training.

\subsection{\label{subsec:training}{Training}}

We train the two \textsc{Genn} models using samples from \textsc{Geant4} simulation. 
We first generate 1,000,000 light sources, uniformly distributed within the detectors' active volume, at vertices ($pos_x$, $pos_y$, $pos_z$). 
From each vertex, 1,000,000 photons are emitted. 
Those photons are transported inside the active volume and a certain number of photons are detected by particular photon detectors. 
For each scintillation vertex, we get an ``image'' of the hit pattern on the photon detectors, which is organized as a 1D-vector. 
We produce the images by simulating the photon transport in ProtoDUNE-like and DUNE-like geometries using the \textsc{LArSoft} toolkit \cite{snider2017larsoft} which uses \textsc{Geant4}. 
Combining the scintillation vertex and the image, we build the training samples in the structure: ($'x':~pos_x,~'y':~pos_y,~'z':~pos_z,~'image':~image$). 
We emphasize that the sequence for elements in the 1D-vector for $image$ is plainly organized by the sequence number of the photon detectors instead of flattening the human-friendly spatial-representative image.

We train the \textsc{Genn} models with a batch size of 4096, using the \textit{Adam} optimizer \cite{kingma2014adam} on the Wilson Cluster\footnote{Wilson Cluster: https://computing.fnal.gov/wilsoncluster/} at Fermilab with two NVIDIA Tesla K40 GPUs. In order to help the model converge quickly and stably, we use a learning rate scheduler to reduce the learning rate during training. The learning rate is initialized as: $lr = 0.0002$, and decays following: $lr = lr \times 0.997^{\rm epoch}$. In order to help the model get out of potential saddle points, we reset the learning rate to its initial value every 1000 epochs. The training procedure is monitored with the metric ``mean absolute error of the validation sample'' to avoid overtraining. The training procedure stops automatically when the monitored metric ``validation sample loss'' stops improving after 500 epochs, which was typically less than 10,000 epochs and within 8 hours.

\section{\label{sec:perf}{Performance evaluation}}

Once the networks were trained, we evaluated the \textsc{Genn} models' feature-learning capability and benchmarked their performance within the \textsc{LArSoft} framework.

\subsection{\label{subsec:feat}{Feature-learning capability}}

A photon detector observes more photons when a light source is closer to it. Hence, the probability of photon being detected by the particular photon detector, or the so-called ``visibility'', is higher. Consequently, the ``visibility'' of a photon detector reaches a peak when the light source moves towards it and reduces when that moves away. In addition, both the \textsc{Genn} prediction and \textsc{Geant4} simulation are expected to see a smooth variation in the total amount of light detected by particular photon detectors when the light source passes by.

\begin{figure*}[htp]
\centering
\includegraphics[width=0.95\textwidth]{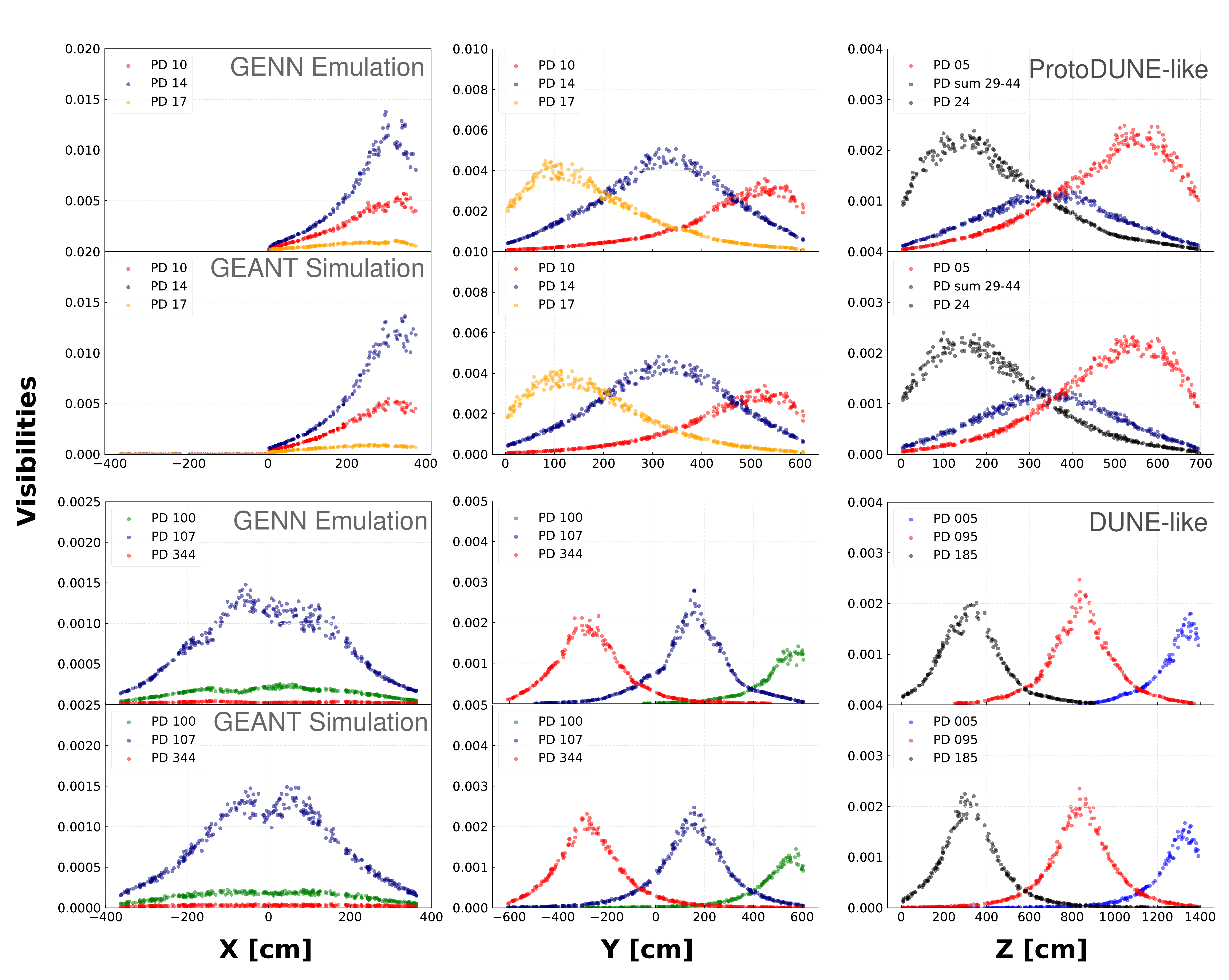}
\caption{Feature-learning capability. In each figure, the $x$ axis is the position of the light sources along specific directions: $x$, $y$, and $z$ direction, while the $y$ axis shows the photon detection probabilities (visibility), on particular photon detectors (PDs). The top two rows show the visibilities in ProtoDUNE-like geometry while the bottom two are for the DUNE-like geometry. The visibilities from \textsc{Genn} prediction are plotted on the first and third rows and those from \textsc{Geant4} simulation are on the second and forth rows. Note: positions of particular PDs and light sources are showed in Table.~\ref{tab:pdpos}.}
\label{fig:behavior}
\end{figure*}

In order to verify the models have learned desired characteristics, we compose three groups of light sources in \textsc{LArSoft}, which are inside detectors' active volume and distributed uniformly along $x$, $y$, and $z$ direction respectively, and predict the ``visibility'' using \textsc{Genn} and \textsc{Geant4}. For each group, we study the visibilities on three photon detectors. The position of the photon detectors and the light sources are shown in Table~\ref{tab:pdpos}. The ``visibilities'', produced by \textsc{Genn} and \textsc{Geant4}, are plotted in Figure~\ref{fig:behavior} for comparison.

\begin{table*}[htp]
\centering
\caption{\label{tab:pdpos} 3D-position of the photon detectors (PDs) being studied. Three groups of light sources are produced and uniformly distributed along $x$, $y$, and $z$ direction.
}
\scriptsize  % \tiny \scriptsize \footnotesize \small \normalsize \large \Large \LARGE \huge \Huge
\begin{tabular*}{\textwidth}{ c @{\extracolsep{\fill}} c c c c}
\hline
\hline
\multicolumn{2}{c}{ProtoDUNE-like} & \multicolumn{2}{c}{DUNE-like} \\
\hline
PD\#  & Position       &  & PD\#    & Position \\
\hline
10    & 363, 568, 347  &      & 100  & ~0.05, ~591, 781  \\
14    & 363, 331, 347  &      & 107  & ~0.05, ~155, 781  \\
17    & 363, ~94, 347  &      & 344  & -0.05, -280, 781  \\
05    & 363, 272, 579  &      & 005  & ~0.05, ~280, 1357 \\
29-44 & 363, 272, 428$\sim$265 &  & 095  & ~0.05, ~280, 843  \\
24    & 363, 272, 115  &      & 185  & ~0.05, ~280, 3169 \\
\hline
\multicolumn{5}{c}{Position of Light Sources} \\
\hline
\multicolumn{2}{c}{-400$\sim$400, 420$\pm$10, 345$\pm$10} & & \multicolumn{2}{c}{-400$\sim$400, 300$\pm$10, 780$\pm$10} \\
\multicolumn{2}{c}{150$\pm$10, 0$\sim$600, 345$\pm$10} & & \multicolumn{2}{c}{150$\pm$10, -600$\sim$600, 780$\pm$10} \\
\multicolumn{2}{c}{150$\pm$10, 420$\pm$10, 0$\sim$700} & & \multicolumn{2}{c}{150$\pm$10, 300$\pm$10, 0$\sim$1400} \\
\hline
\hline
\end{tabular*}
\end{table*}

Figure~\ref{fig:behavior} shows the ``visibilities'' of the photon detectors change smoothly when the light source ``moves'', which matches the general trends of what is produced from full \textsc{Geant4} simulation.  This test indicates that the \textsc{Genn} models are able to learn features from the dataset and interpolate smoothly, instead of only memorizing the training samples. Figure~\ref{fig:behavior} indicates \textsc{Genn} model can generalize what it has learned and predict probabilities of photon detected by particular photon detectors with precision similar to \textsc{Geant4}'s.

\subsection{\label{subsec:acc}{Prediction precision}}

We evaluate the precision of the \textsc{Genn} prediction in the \textsc{LArSoft} framework. After being trained, the models and weights are frozen to computable graphs which are loaded by the \textsc{TensorFlow} C++ API. We simulate the transport of samples representative of the physics of interest to neutrino experiments in \textsc{Geant4}, where we use 200~MeV monoenergetic muons starting from a specific position in the ProtoDUNE-like geometry and supernova neutrinos uniformly distributed in the DUNE-like geometry, and record the points of energy depositions (scintillation vertices). We construct light sources at each scintillation vertex, where the number of photons emitted is calculated based on the deposited energy. Finally, we use \textsc{Genn} models to predict and \textsc{Geant4} simulation to produce photon detection probabilities with respect to individual light sources. We then compare the performance of \textsc{Genn} prediction to \textsc{Geant4} simulation at the same level of detail. 

\begin{figure*}[htp]
\centering
\includegraphics[width=0.9\textwidth]{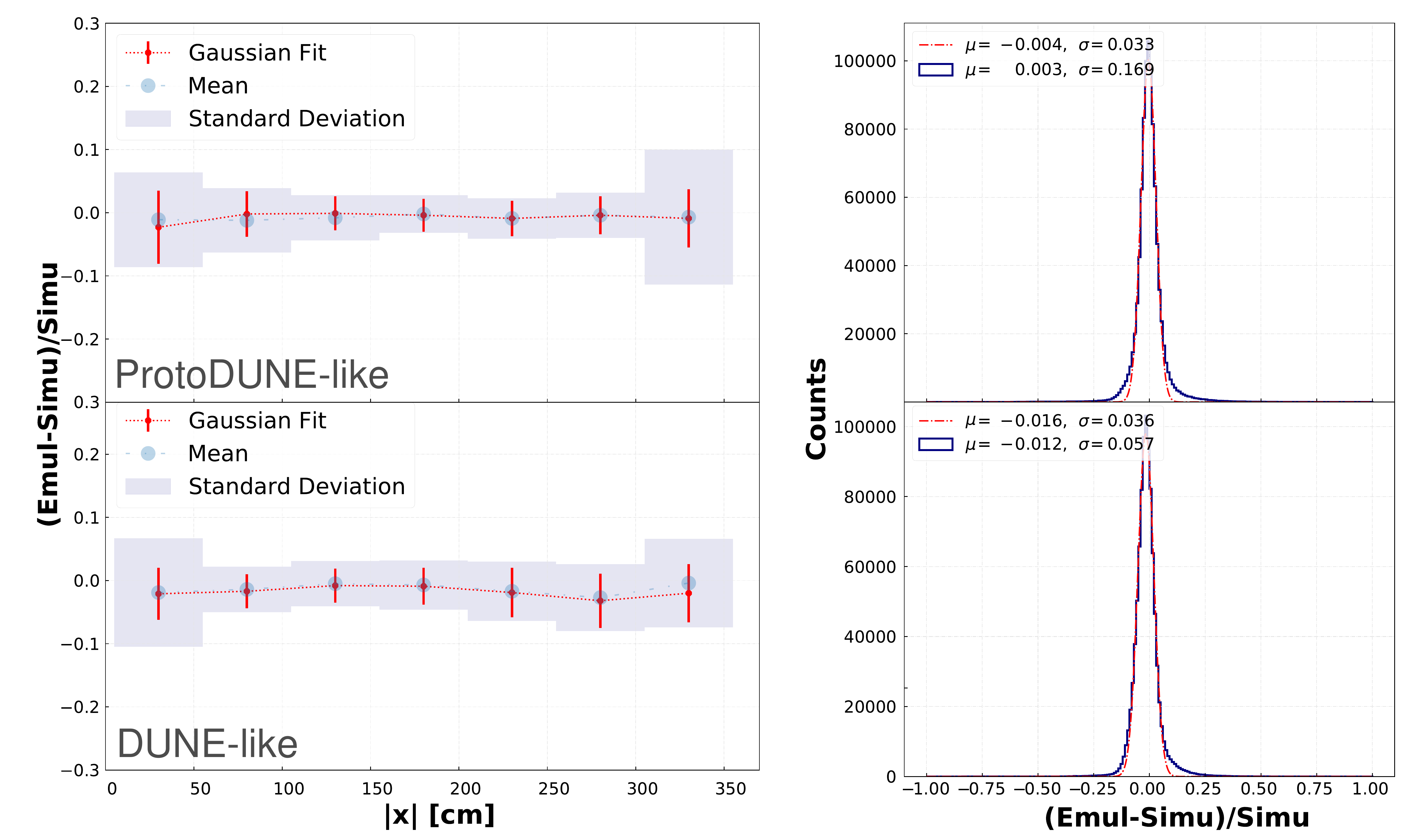}
\caption{Discrepancy of total detected photons per scintillation vertex. Left: discrepancies variation along the $x$-axis. Right: overall discrepancies, where the red dashed line is a Gaussian fit. Comparison for ProtoDUNE-like geometry is on the top and DUNE-like geometry on the bottom.}
\label{fig:int}
\end{figure*}

The number of total detected photons, used for event energy reconstruction, is critical information we want to extract from the photon detection system. We obtain the number of total detected photons from a scintillation vertex by summing up the number of detected photons on each photon detector, and calculate the relative difference between the two approaches for each vertex. The distribution of these relative differences is shown in Figure~\ref{fig:int}, along with a Gaussian fit to the peak.

It shows that the \textsc{Genn} model for the ProtoDUNE-like geometry reproduces the total number of photons detected from each scintillation vertex with a resolution of 3.3\%, and a mean 0.4\% less than the \textsc{Geant4} simulation, while the model for DUNE-like geometry has a resolution of 3.6\% and on average produces 1.6\% fewer photons. % The number of scintillation vertices outside the 1$\sigma$ region of the Gaussian fit is 50.09\% in ProtoDUNE-like geometry and 56.14\% in DUNE-like geometry.
The fraction of scintillation vertices whose deviation is larger than 10\% is 8.95\% in ProtoDUNE-like geometry and 7.69\% in DUNE-like geometry. %\TK{Fraction outside 10\% I think will be more informative.}
It indicates the 1-neuron $vis_{\rm int}$-layer is able to accurately normalize the number of total detected photons.

\begin{figure*}[htp]
\centering
\includegraphics[width=0.9\textwidth]{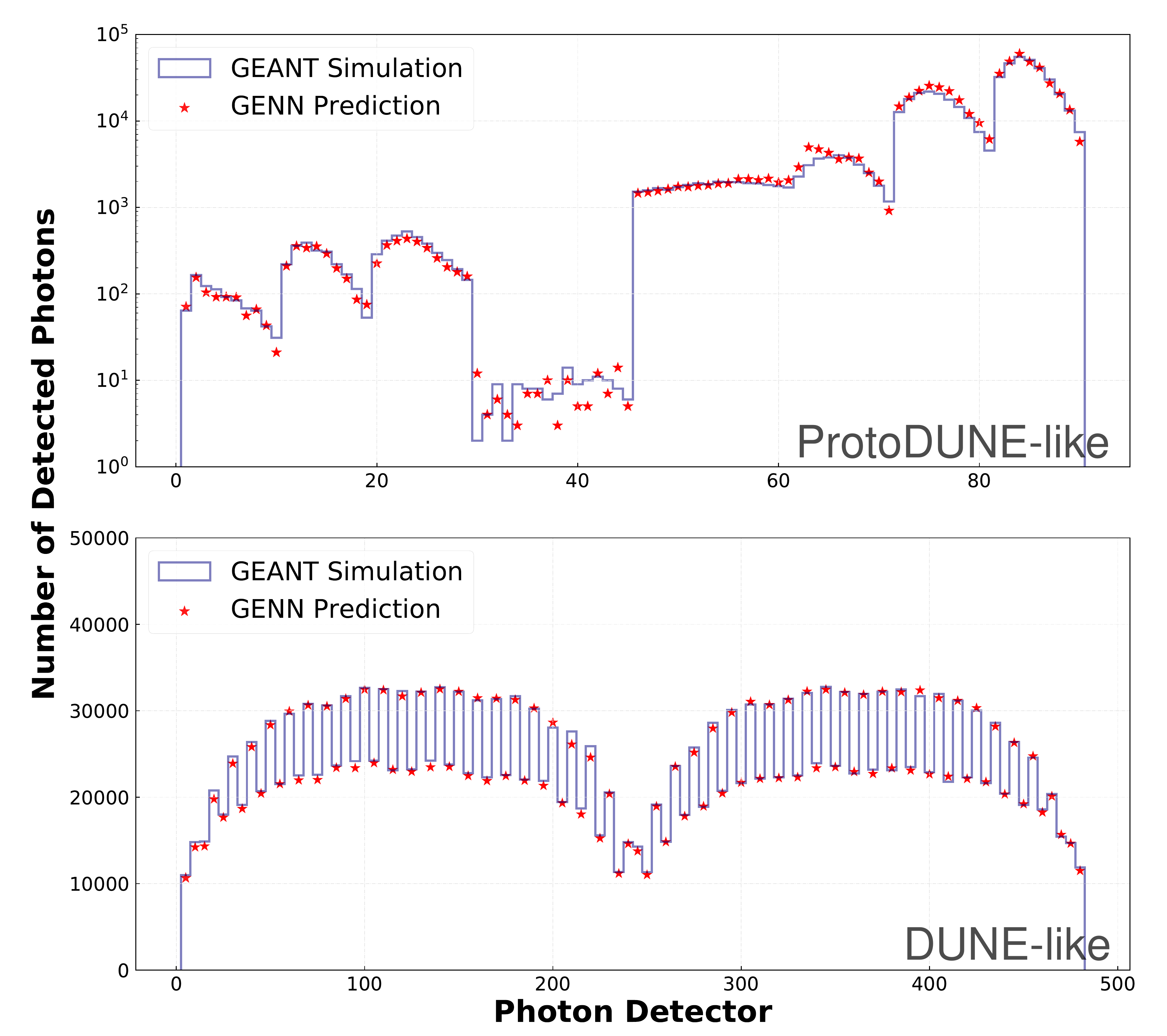}
\caption{Distribution of the detected photons on photon detection system. Top: the sum of detected photons on photon detectors from 200~MeV monoenergetic muon events starting from one specific position in the ProtoDUNE-like geometry. The response is much lower around photon detector \#40 is caused by the fact that those photon detectors are the segmented ones which are much smaller than the bar-like light collectors. Bottom: the sum of detected photons on photon detectors from supernova neutrinos uniformly distributed in the DUNE-like geometry.}
\label{fig:pat}
\end{figure*}

Besides the number of total detected photons, we need the model to predict the photon distribution on photon detectors, which is valuable for position reconstruction of the scintillation vertex. Figure~\ref{fig:pat} shows the number of photons detected by each photon detector for both the \textsc{Geant4} simulation and the \textsc{Genn} prediction. Both methods give similar photon distributions for both photon detection systems. Although the photon detection system deployed in ProtoDUNE-like geometry contains two different types of photon detectors, the \textsc{Genn} model can learn features from training samples, and gives a reasonable prediction, showing $\mathrm{OuterProduct}$-layer can be well trained with the 1D data structure, even in a complex case.

Using the 1D data structure also allows us to calculate some statistical metrics, such as Chi-square distance, Euclidean distance, JS divergence, and MAE, for similarity measures. It also allows us to quantitatively compare the performance between different emulation or simulation approaches, but this is beyond the scope of this paper.

\subsection{\label{subsec:inf}{Inference performance}}

One of the goals for the \textsc{Genn} model is to predict the photon detection probabilities at high speed using a CPU. To measure the computing performance of \textsc{Genn} models, we use the method mentioned in Section.~\ref{subsec:acc} to build light sources along the track of ionizing particles: 200-MeV monoenergetic muons, 2-GeV monoenergetic electrons, and supernova neutrinos. We load those light sources in \textsc{LArSoft} to run \textsc{Genn} prediction and \textsc{Geant4} simulation for benchmarking. In Table~\ref{tab:inf}, we present the CPU time per event, per image, and per photon to compare the computing performance between the two approaches. Thanks to the $\mathrm{OuterProduct}$-layer and the $\mathrm{Multiply}$-layer, the \textsc{Genn} model introduced in this paper is lightweight, and the \textsc{Genn} prediction speed is 20 to 50 times faster than \textsc{Geant4} simulation while keeping same level of detail and precision.

\begin{table*}[htp]
\centering
\scriptsize  % \tiny \scriptsize \footnotesize \small \normalsize \large \Large \LARGE \huge \Huge
\caption{\label{tab:inf}CPU time for \textsc{Geant4} simulation and \textsc{Genn} prediction. Events used for the test: 200~MeV monoenergetic $\mu$, 2~GeV monoenergetic $e^-$, and Supernova $\nu_e$. All benchmarks for CPU time are performed on DUNEGPVM at Fermilab where 2.4 GHz 4-core Intel$^\circledR$ Core$^\text{TM}$ Processor (Broadwell) are deployed.}
\begin{tabular}{l l c c c c c c c}
\hline
\hline
&                            & \multicolumn{3}{c}{\textsc{Geant4} Simulation} & &  \multicolumn{3}{c}{\textsc{Genn} Prediction} \\
&                    & (s/event)              & (ms/image)       & ($\mu$s/photon)  &   & (s/event) & (ms/image)     &  ($\mu$s/photon) \\
\hline
\multirow{2}{*}{ProtoDUNE-like} & $\mu$   & 3.3$\pm$0.5 & 3.71$\pm$0.51   & 34.32$\pm$4.72  & & 0.14$\pm$0.04 & 0.15$\pm$0.04  & 1.43$\pm$0.40  \\
                                & $e^-$   & 61.1$\pm$0.5 & 3.24$\pm$0.03   & 44.54$\pm$0.39  &  & 2.83$\pm$0.10 & 0.15$\pm$0.01  & 2.06$\pm$0.08  \\
\hline
\multirow{2}{*}{DUNE-like}      & $\nu_e$ & 127.4$\pm$2.9 & 10.12$\pm$0.23  & 85.05$\pm$1.92  & & 2.65$\pm$0.15 & 0.21$\pm$0.01  & 1.77$\pm$0.10  \\
                                & $e^-$   & 103.6$\pm$2.9 & 6.76$\pm$0.19   & 75.62$\pm$2.13  & & 2.74$\pm$0.09 & 0.18$\pm$0.01  & 2.00$\pm$0.07  \\
\hline
\hline
\end{tabular}
\end{table*}

%\begin{table*}[htp]
%\begin{ruledtabular}
%\begin{tabular}{l l c c c c}
%&                            & \multicolumn{2}{c}{\textsc{Geant4} Simulation} & \multicolumn{2}{c}{\textsc{Genn} Prediction} \\
%&                            & (ms/image)       & ($\mu$s/photon)             & (ms/image)     &  ($\mu$s/photon) \\
%\hline
%\multirow{2}{*}{ProtoDUNE-like} & $\mu$   & 3.71$\pm$0.51   & 34.32$\pm$4.72  & 0.15$\pm$0.04  & 1.43$\pm$0.40  \\
%                                & $e^-$   & 3.24$\pm$0.03   & 44.54$\pm$0.39  & 0.15$\pm$0.01  & 2.06$\pm$0.08  \\
%\hline
%\multirow{2}{*}{DUNE-like}      & $\nu_e$ & 10.12$\pm$0.23  & 85.05$\pm$1.92  & 0.19$\pm$0.01  & 1.77$\pm$0.10  \\
%                                & $e^-$   & 6.76$\pm$0.19   & 75.62$\pm$2.13  & 0.18$\pm$0.01  & 2.00$\pm$0.07  \\
%\end{tabular}
%\end{ruledtabular}
%\caption{\label{tab:inf}CPU time for \textsc{Geant4} simulation and \textsc{Genn} prediction. Events used for the test are: 200~MeV monoenergetic $\mu$, 2~GeV monoenergetic $e^-$, and Supernova $\nu_e$ events. All benchmarks for CPU time are performed on DUNEGPVM at Fermilab where 2.4 GHz 4-core Intel$^\circledR$ Core$^\text{TM}$ Processor (Broadwell) are deployed.}
%\end{table*}

The important information we extract from the test is: \textsc{Genn} prediction run time only depends on the complexity of the models, while \textsc{Geant4} simulation run time increases significantly with detector volume. The total CPU time for an event will be determined by the number of photons and detector volume for \textsc{Geant4} simulation, and only by the number of energy deposition vertices along the particle's track for \textsc{Genn} prediction. Since the granularity of the energy depositions along the track can be adjusted, the \textsc{Genn} prediction can be accelerated even further at the expense of less detail on the particle tracks. 

Another key advantage of the \textsc{Genn} model is that the model inference requires relatively little memory. The samples for ProtoDUNE-like and DUNE-like geometries show the required memory for the model inference is around 15\% of the \textsc{Geant4} simulation. Further, this memory use is not directly correlated to the volume of the detectors, unlike using lookup libraries where the available memory on the machine limits the potential granularity (and hence precision).

\subsection{\label{subsec:sca}{Scalability}}

\begin{figure*}[htp]
\centering
\includegraphics[width=0.9\textwidth]{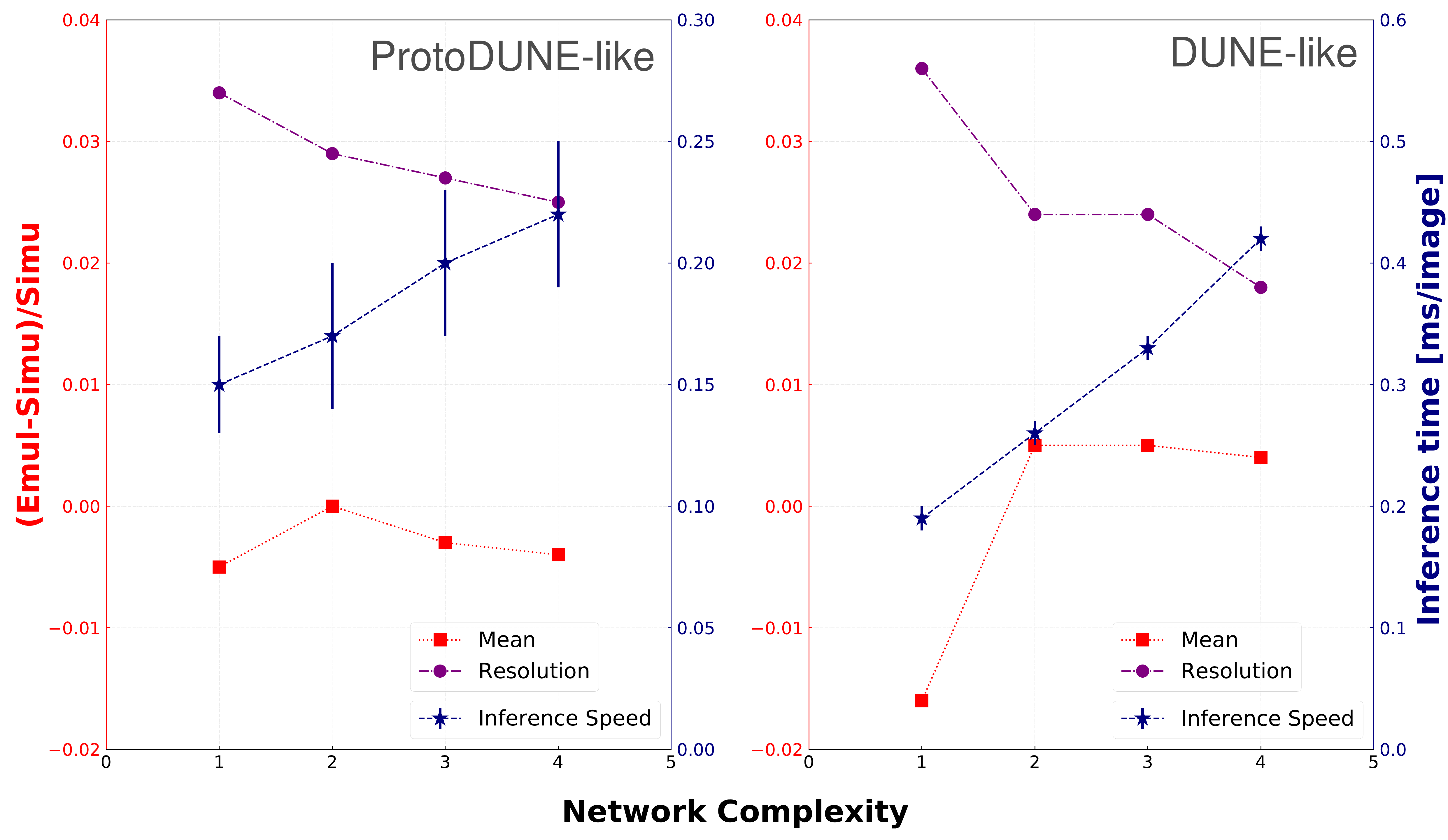}
\caption{Precision ($y$ axis on the left), inference time ($y$ axis on the right), and neural network complexity.}
\label{fig:com}
\end{figure*}

While the \textsc{Genn} model already looks promising, it is possible to add more ``optional'' layers for higher precision, with a trade-off that the inference time increases. We conduct an empirical study to quantify both inference time and precision at different levels of network complexity. In particular, we construct \textsc{Genn} models at four levels of complexity by introducing 1 to 4 additional dense layers. The comparison of the precision and inference performance between the four \textsc{Genn} models is showed in Figure~\ref{fig:com}. Our observation indicates a balance between the inference speed and precision must be found when choosing the depth of the network.

As the \textsc{Genn} model prediction run time depends on the network complexity, we benchmark the inference performance for photon detection systems with different number of photon detectors. The metrics in Table~\ref{tab:spd} reveal the inference time increases linearly with the number of photon detectors. The \textsc{Genn} model still outperforms \textsc{Geant4} simulation, at least, by a factor of 10 even when 1,920 photon detectors are deployed in DUNE-like geometry.

\begin{table*}[htp]
\centering
\scriptsize  % \tiny \scriptsize \footnotesize \small \normalsize \large \Large \LARGE \huge \Huge
\caption{\label{tab:spd}Inference time (CPU) for difference photon detection systems.}
\begin{tabular}{l c c c c  c c c c}
\hline
\hline
Geometry        & \multicolumn{2}{c}{ProtoDUNE-like}  &~~~   & \multicolumn{5}{c}{DUNE-like}                                    \\
Num. of PDs     & 60            & 90                  &~~~   & 120           &  240          & 480           & 960           & 1920          \\
\hline
Time (ms/image) & 0.11$\pm0.02$ & 0.15$\pm0.03$      &~~~  & 0.14$\pm0.01$ & 0.17$\pm0.01$ & 0.21$\pm0.01$ & 0.41$\pm0.01$ & 0.87$\pm0.01$ \\
\hline
\hline
\end{tabular}
\end{table*}

\section{\label{sec:con}{Conclusion and outlook}}

In this paper we present a novel generative model for photon detection probability prediction. This \textsc{Genn} model uses an $\mathrm{OuterProduct}$-layer to predict the photon distribution on photon detectors and a single-neuron layer to normalize the photon intensity. This architecture realizes the ``deconvolution'' from the scintillation vertex to the photon detection probability and makes the model prediction precise and fast. We propose an optimal loss-function using a variational KL-divergence function for the model training, which is able to be generalized for common 1D generative models.

The model built for ProtoDUNE-like photon detection system demonstrates that shallow neural networks are able to learn features from training samples represented by 1D data structures, even for complex photon detection systems. The sample for DUNE-like photon detection system indicates the \textsc{Genn} model gives fast and precise prediction of photon detection probability using limited memory, showing it can be a powerful new tool to bypass the full \textsc{Geant4} simulation in a production environment, especially for LAr detectors with huge volumes, such as the DUNE far detector. The successful application to the ProtoDUNE-like and DUNE-like geometries shows this general \textsc{Genn} architecture is easy to generalize for different photon detection systems.

Our future attention will focus on incorporating the most recent cutting-edge neural network architecture to improve the prediction precision and the inference speed. We will also develop new photon simulation strategies when using \textsc{Genn} models, guided by the physics goals for dark matter experiments or neutrino experiments.

While our primary effort will be to keep improving this model application for photon detection probability prediction in large scale LAr detector, this \textsc{Genn} model and the $D_{\rm vKL}$-loss-function are quite general and may be applied in other contexts where the training data can be represented as a 1D vector and the \textsc{\textsc{Gan}} framework might not work efficiently \cite{zaheer2018gan}.

\acknowledgments

The authors thank S.~Alonso Monsalve, E.~Church, T.~Junk, A.~M.~Szelc, L.~H.~Whitehead, and T.~Yang for helpful comments and discussion. This work is supported by the U.S. Department of Energy through the Early Career Award Program. This manuscript has been authored by Fermi Research Alliance, LLC under Contract No. DE-AC02-07CH11359 with the U.S. DOE, Office of Science, Office of High Energy Physics and prepared using the resources of the Fermi National Accelerator Laboratory.


\bibliographystyle{JHEP}
\bibliography{main}
%\begin{thebibliography}{99}
%\end{thebibliography}
\end{document}